\documentclass[12pt]{book}

\usepackage[dvips]{graphicx,color}
\usepackage{makeidx,tsukuba}

\makeauthorindex
\makeindex

\begin{document}

\BookTitle{\itshape The 28th International Cosmic Ray Conference}
\CopyRight{\copyright 2003 by Universal Academy Press, Inc.}
\pagenumbering{arabic}

\chapter{The GZK Feature in the Spectrum of UHECRs:\\ 
What is it Telling Us?}

\author{%
Daniel De Marco,$^1$ Pasquale Blasi,$^2$ Angela V.\ Olinto,$^{3,4}$\\
{\it 
(1) Universit\`a di Genova, Dipartimento di Fisica \& INFN Sezione di
Genova, Genova, ITALY\\ 
(2) INAF/Osservatorio Astrofisico di Arcetri, Firenze, ITALY\\ 
(3) Center for Cosmological Physics, The University of Chicago, Chicago, USA\\
(4) Department of Astronomy \& Astrophysics,  \& Enrico Fermi Institute,
The University of Chicago, Chicago,  USA
}
}

\section*{Abstract} 
We developed a numerical simulation of the propagation of UHECR in the
Cosmic Microwave Background (CMB) and we used it to determine the significance
of the GZK feature in the spectrum of UHECR measured by AGASA and HiRes.
We find that these two experiments are best fit by two different
injection spectra in the region below $10^{20}$~eV and that the error
bars around the GZK feature are dominated by fluctuations which leave a
determination of the GZK feature not attainable at present. In addition
the comparison of the spectra of AGASA and HiRes suggests the presence
of about a 30\% systematic errors in the relative energy determination
of the two experiments. Correcting for these systematics, the two
experiments are brought in agreement at energies below $10^{20}$~eV and,
in this region, are best fit by an injection spectrum with spectral
index 2.5-2.6. In the high energy region (above $10^{20}$~eV) the two
experiments maintain their disagreement, but only at the 2$\sigma$
level. Our results clearly show the need for much larger experiments
such as Auger, EUSO, and OWL, that can increase the number of detected
events by one or two orders of magnitude.

\section{Introduction}
Astrophysical proton sources distributed homogeneously in the universe
produce a feature in the energy spectrum due to the production of pions
off the CMB. This feature, consisting of a
rather sharp suppression of the flux, occurs at energies above $E_{\rm
GZK} \sim7 \times 10^{19}$~eV, and it is now known as the GZK
cutoff. Almost forty years after this prediction it is not
yet clear if this effect is observed or not due to the discrepancy
between the results of the two largest experiments measuring the
spectrum of Ultra High Energy Cosmic Rays (UHECRs). AGASA[1] reports a
higher number of events above $E_{\rm GZK}$ than expected while HiRes[2]
reports a flux consistent with the GZK feature.  Here we report on a
detailed investigation[4] of the statistical
significance of this discrepancy as well as the significance of the
presence or absence of the GZK feature in the data. We find that neither
experiment has the necessary statistics to establish if the spectrum of
UHECRs has a GZK feature.  In addition, a systematic error in the energy
determination of the two experiments seems to be required in order to
make the two sets of observations compatible in the low energy range,
$10^{18.5}-10^{19.6}$ eV, where enough events have been detected to make
the measurements reliable. Taking into account the systematics, the two
experiments predict compatible fluxes at energies below $E_{\rm GZK}$
and at energies above $E_{\rm GZK}$ the fluxes are within $\sim 2\sigma$
of each other. 



\begin{figure}
\begin{center}
\includegraphics[width=0.4\textwidth,height=0.25\textwidth]{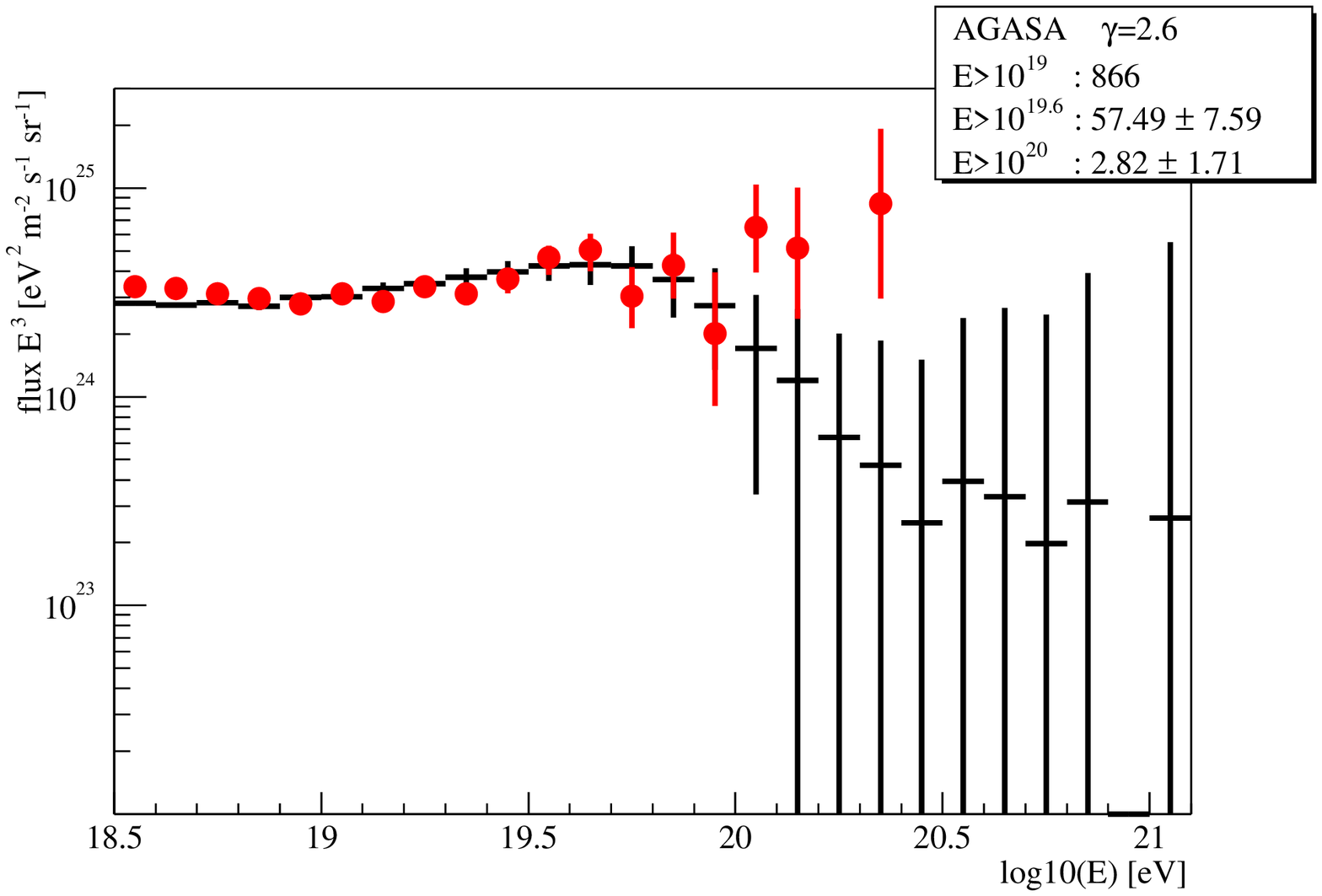}
\includegraphics[width=0.4\textwidth,height=0.25\textwidth]{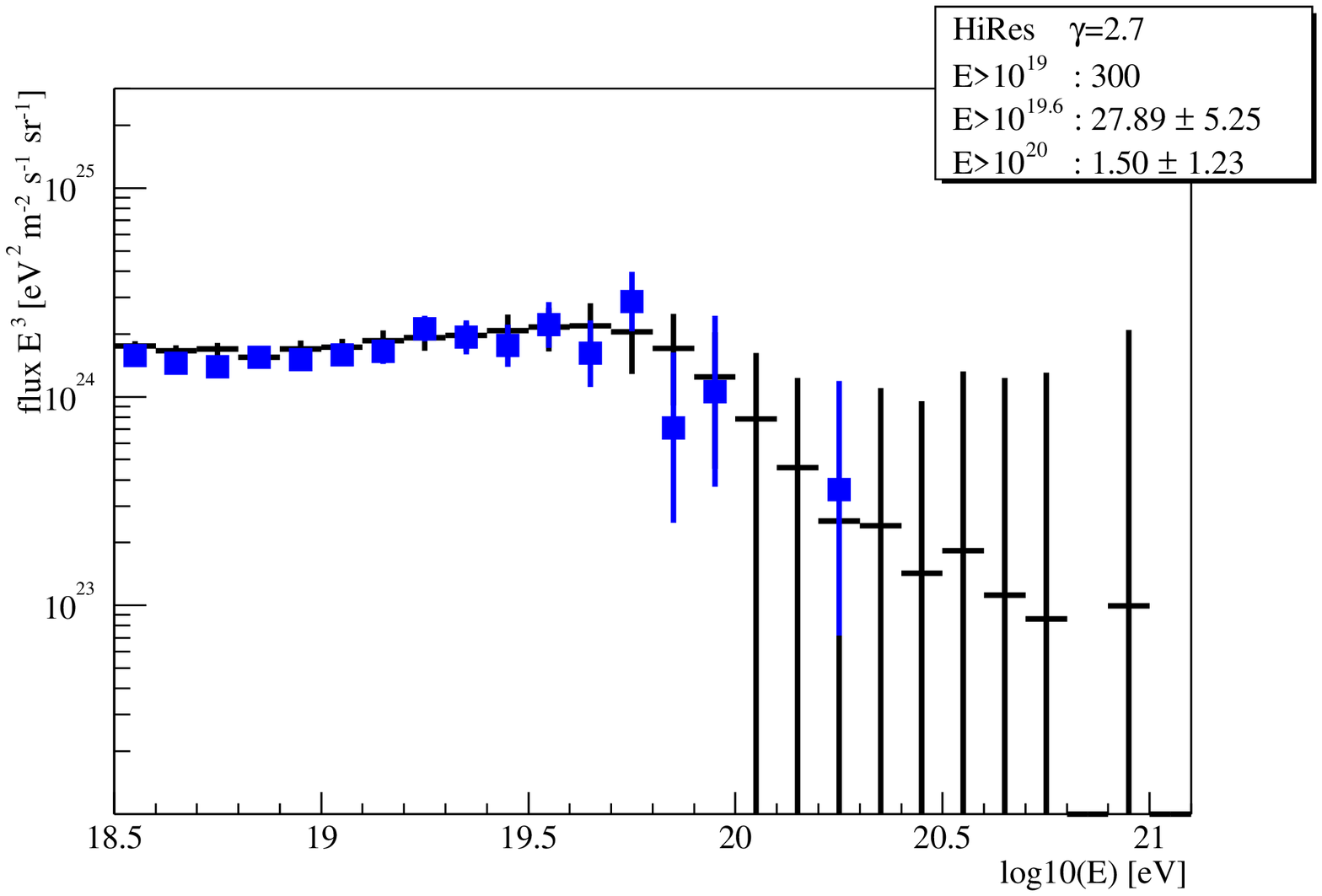}
\end{center}
\vspace{-1.5pc}
\caption{Simulations of AGASA and HiRes statistics.
crosses with error bars: simulations, grey points: experimental data.}
\vspace{-1.5pc}
\end{figure}

\section{The cosmic rays propagation code}
We assume that UHECRs are protons injected in extragalactic sources with
a power-law spectrum with slope $\gamma$ and an exponential cutoff at
$E_{\rm max}=10^{21.5}\,{\rm eV}$. Based on the results of [3] we assume
a spatially uniform distribution of sources and do not take into account
luminosity evolution in order to avoid the introduction of additional
parameters.
We simulate the propagation of protons from source to observer by
including the photo-pion production, pair production, and adiabatic
energy losses due to the expansion of the universe. We calculate the
pair production and adiabatic losses using the continuous energy loss
approximation while for the photo-pion production we use a montecarlo
approach given the high inelasticity of this interaction [4]. 
We propagate particles until the statistics of events detected above
some energy reproduces the experimental numbers. By normalizing the
simulated flux by the number of events above an energy where experiments
have high statistics, we can then ask what are the fluctuations in
numbers of events above a higher energy where experimental results are
sparse. 
We study the spectrum above $10^{18.5}$ eV where the flux is supposed to
be dominated by extragalactic sources. For this energy range, we focus
on the experiments that have the best statistics: AGASA and HiResI. For
AGASA data, the simulation is stopped when the number of events above
$E_{th} = 10^{19}$ eV equals 866.  For HiRes this number is 300 (but
needs to be corrected for exposure). We assume for both experiments a
statistical error in the energy determination of 30\%.

\section{AGASA versus HiResI}
The two largest experiments that measured the flux of UHECRs, AGASA and
HiRes, report apparently conflicting results. 
We apply our simulations to the statistics of these experiments in order
to understand whether the discrepancy is statistically significant and
whether the GZK feature has indeed been detected in the cosmic ray
spectrum. In order to do this, we run 400 realizations of the AGASA and
HiRes event statistics for various slopes of the injection spectrum
and using the $\chi^2$ indicator
we find the best fit for each data set. Taking into account the data at
energies above $10^{18.5}$~eV, the best fit spectra are $E^{-2.8}$ for
AGASA and $E^{-2.6}$ for HiRes. Raising the threshold to $10^{19}$~eV
the best fit spectra become $E^{-2.6}$ for AGASA and between $E^{-2.7}$
and $E^{-2.8}$ for HiRes. 
In order to quantify the significance of the detection or lack of
the GZK flux suppression, we count the mean number of events above
$10^{20}$~eV and we compare it to the observed number. For the best fit
injection spectra we find a discrepancy in the range
$2.4-2.8\sigma$ for AGASA and a complete agreement between
simulation and observed data for HiRes. In this comparison $\sigma$ is
simply the error on the observed number of events above $10^{20}$~eV.
Taking into account in this comparison also the theoretical error bars 
makes the significance of the presence or absence of the GZK feature
much weaker, in the range $2.1-2.5\sigma$.
A more graphical representation of the uncertainties involved is
displayed in Fig.~1. 
The large error bars 
generated by our simulations at the high energy end of the spectrum are
mainly due to the stochastic nature of the process of photo-pion
production. The large fluctuations are unavoidable with the extremely
small statistics available with present experiments. On the other hand,
the error bars at lower energies are minuscule, so that the two data
sets (AGASA and HiResI) cannot be considered to be two different
realizations of the same phenomenon. Instead, systematic errors in at
least one if not both experiments are needed to explain the
discrepancies at lower energies where  the two spectra, when multiplied
by $E^3$, are systematically shifted by about a factor of two.
This shift suggests that there may be a systematic error either in the
energy or the flux determination of at least one of the two experiments.
A systematic error of $\sim 15\%$ in the energy determination is well
within the limits that are allowed by the analysis of systematic errors
carried out by both collaborations [1,2], so we split the
energy gap by assuming that the two experiments have a 15\% shift in the
energy determination, but in opposite directions.
For AGASA, the best fit injection spectrum becomes $E^{-2.5}$ above
$10^{19}$~eV and $E^{-2.6}$ above $10^{19.6}$~eV. For HiRes,
the best fit injection spectrum is $E^{-2.6}$ for the whole set of data,
independent of the threshold. It is interesting to note that the best fit
injection spectra as derived for each experiment independently coincides
for the corrected data unlike the uncorrected case. This suggests that
combined systematic errors in the energy determination at the $\sim$
30\% level may in fact be present.

Comparing the observed number of events above $10^{20}$~eV with the
simulated one as in the case without systematics we find that while
HiRes remains in agreement with the prediction of a GZK feature, the
AGASA data seem to depart from such prediction, but only at the level of
$\sim 1.8\sigma$. Taking also into account the theoretical error bars
the discrepancy level drops further to $\sim 1.5\sigma$.
In Fig.~2, we plot the simulated spectra for injection spectrum
$E^{-2.6}$ and compare them to observations of AGASA and HiRes.

\begin{figure}
\begin{center}
\includegraphics[width=0.4\textwidth,height=0.25\textwidth]{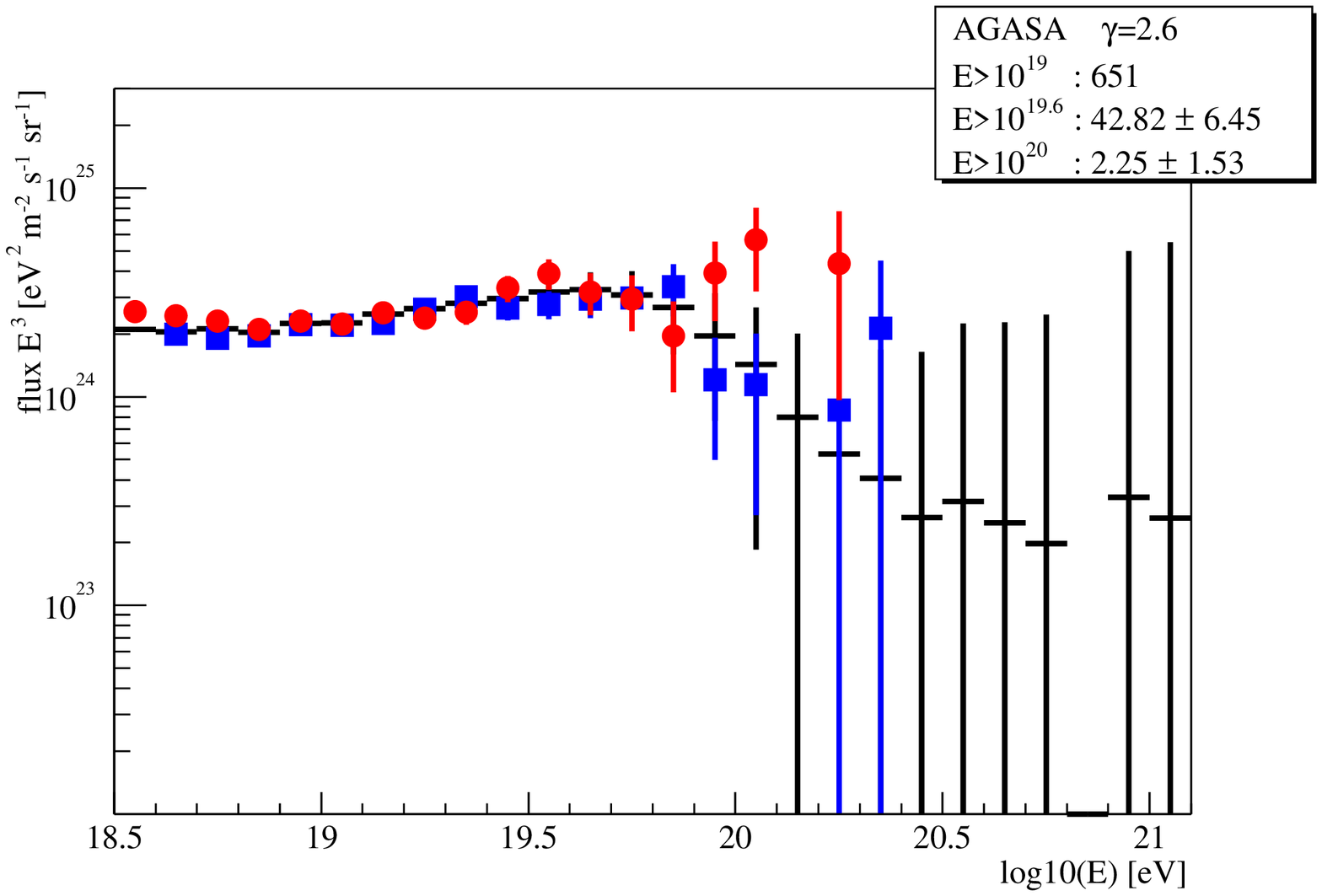}
\includegraphics[width=0.4\textwidth,height=0.25\textwidth]{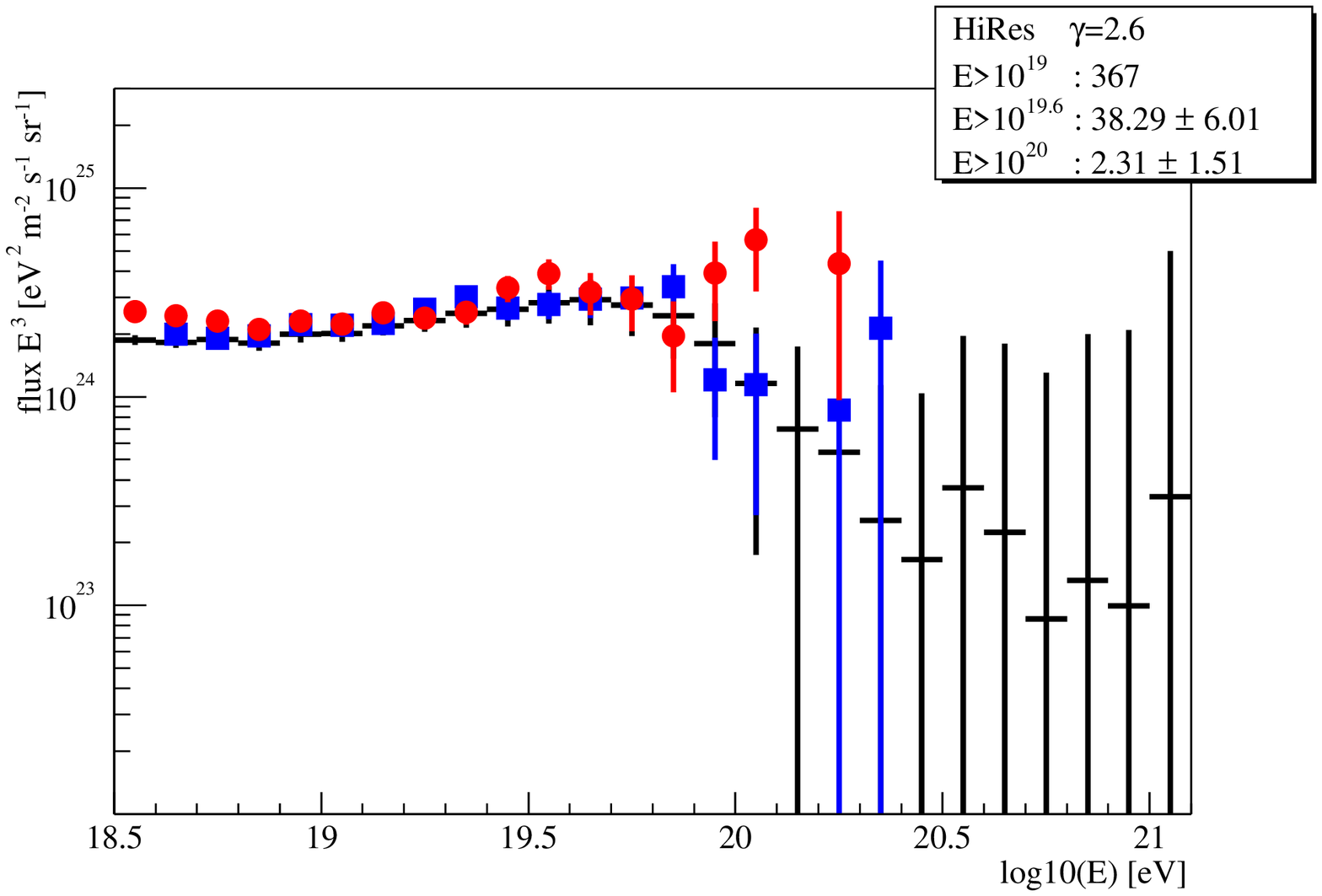}
\end{center}
\vspace{-1.5pc}
\caption{Simulated spectra for the best fit injection spectrum with
$\gamma=2.6$. The shifted data for AGASA (grey circles) and  HiResI
(dark squares) are shown in both panels.}
\vspace{-1.5pc}
\end{figure}

\section{Conclusions}
We compared the spectra obtained from AGASA and HiResI and we found a
systematic shift in the flux, which may be interpreted as a
systematic error in the relative energy determination of about $30\%$.
After the correction for this systematics the two experiment are
basically in agreement.

We considered the statistical significance of these spectra and we found
that with the low statistical significance of either the excess flux
seen by AGASA or the discrepancies between AGASA and HiResI, it is
inaccurate to claim either the detection of the GZK feature or the
extension of the UHECR spectrum beyond $E_{\rm GZK}$ at this point in
time. A new generation of experiments is needed to finally give a clear
answer to this question. The simulated spectra for Auger and EUSO show
that the energy region where statistical fluctuations dominate the
spectrum is moved to $\sim 10^{20.6}$~eV for Auger, allowing a clear
identification of the GZK feature. The fluctuations dominated region
stands beyond $10^{21}$ eV for EUSO [4].

\section{References}
\re
1.\ Hayashida N. et al., astro-ph/0008102 and ref. therein
\re
2.\ Abu-Zayyad T. et al., astro-ph/0008301, astro-ph/0008243 and ref. therein
\re
3.\ Blanton M., Blasi P., Olinto A.V. 2001, Astropart. Phys. 15, 275
\re
4.\ De Marco D., Blasi P., Olinto A.V. astro-ph/0301497, Astropart.
Phys. in press

\endofpaper
\end{document}